\documentclass[12pt]{iopart}

\usepackage{iopams}  
\usepackage{epsfig}
\usepackage{graphicx}

\begin{document}

\title[Influence of  J$_H$ and SOC on spectral properties of fcc-Ce: DFT+DMFT study.]{Influence of  Hund's exchange and spin-orbit coupling on spectral properties of fcc-Ce: DFT+DMFT study}

\author {A.~O.~Shorikov, V.~I.~Anisimov}

\address{Institute of Metal Physics, Russian Academy of Sciences, 620990 Yekaterinburg, Russia\\
Urals State Technical University, 620002, Mira st. 19, Yekaterinburg, Russia }

\begin {abstract}

Spectral properties of fcc-Ce have been calculated in frames of modern DFT+DMFT method with  Hybridization expansion CT-QMC solver. The influence of  Hund's exchange  and spin-orbit coupling (SOC) on spectral properties of Ce were investigated. SOC is responsible for the shape of spectra near the Fermi level and Hund's exchange interaction doesn't change the obtained spectra and can be neglected. 

\end {abstract}


\maketitle

Rich phase diagram of cerium draws a lot  of attention of researchers. The isostructural $\alpha-\gamma$ transition in Ce is one of the classical problems in the modern solid states physics. In the low temperature $\alpha$-phase (up to T$\sim 100$~K at normal conditions, or until T$\sim 300$~K for P=1~GPa) Ce behaves like a Pauli paramagnet, while in the high temperature $\gamma$-phase the susceptibility approximately follows a Curie-Weiss law.~\cite{Koskenmaki78} The transition is accompanied by a drastic volume collapse (9-15\%)~\cite{Koskenmaki78} and dramatic changes of the electronic spectra.~\cite{Liu92}

Many models was proposed to describe $\alpha$-$\gamma$ transition in Ce. One of the first supposed that localized 4f electrons were transfered to the $spd-$(valence) band state, losing  their local moments.~\cite{Pauling47}.  This model contradicts to later experimental results that show that the number of 4f electrons is almost unchanged during the transition.~\cite{Gustafson69}  Basing of these data a Mott-like picture was proposed, where the transition, which affects the degree of 4f electron localization, occurs due to change of the ratio of on-site $f-f$ Coulomb interaction ($U$) to kinetic energy.~\cite{Johansson74}  Further neutron experiments~\cite{Murani05} confirmed that the Kondo volume collapse model since the Ce-4f electrons remain localized during $\alpha$-$\gamma$ transition. Phase transition thermodynamics in Ce were investigated in frames DFT. It was shown that the critical point cannot be reproduced within  classical DFT approach.~\cite{Wang00} Free energy as a function of temperature was calculated usinf LDA and SIC-LSD methods and the role of entropy in $\alpha$-$\gamma$ transition was emphasized ~\cite{Luders05}~\cite{Amadon06}. Nevertheless DFT fails to reproduce Ce spectral properties as quasiparticle peak near Fermi level and Hubbard bands. The state-of-the-art DFT+DMFT method~\cite{LDA+DMFT} was applied to investigate Ce problem. Perturbation theory was used to investigate spectral properties of $\alpha$- and $\gamma$-Ce within NCA approximation~\cite{Zolfl01} and optical properties  within one-crossing approximation (OCA).~\cite{Haule05} LDA+DMFT method with HF-QMC solver was used to understand the mechanism of $\alpha$-$\gamma$ transition.~\cite{Held01}. Magnetic susceptibility of Ce was calculated using DFT+DMFT method with CT-QMC impurity solver.~\cite{Streltsov12}. Both quasiparticle peak and Hubbard bands were reproduced in DMFT calculation but complicate shape of spectra near Fermi level were committed. Splitting of peak near Fermi lever could originate from crystal field splitting or strong SOC renormalized due to correlations. In this work the influence of Hund's exchange and SOC on spectral properties of Ce were investigated in frame of DFT+DMFT. 

\begin {figure}
\includegraphics [width=0.45\textwidth]{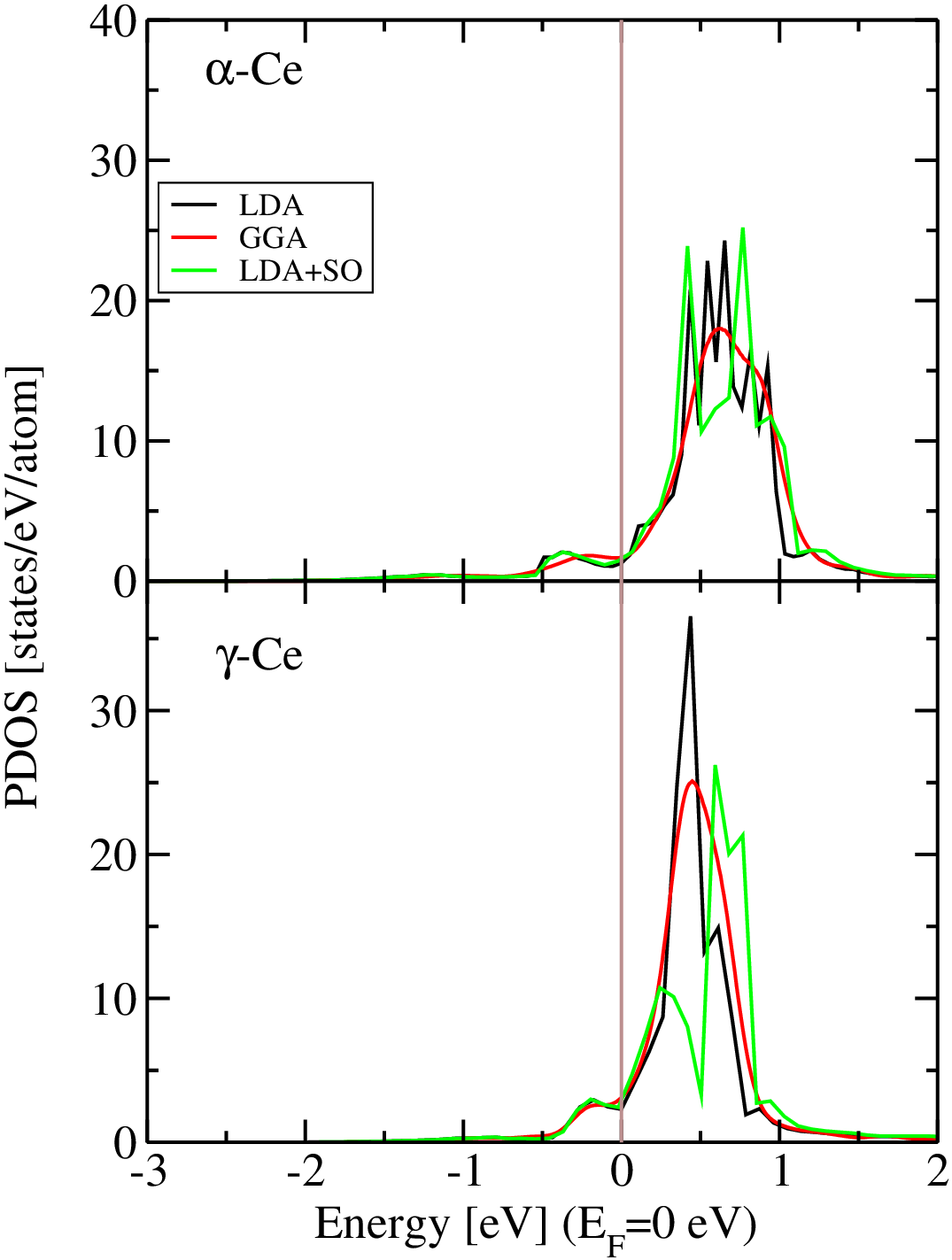}
\caption {DOS of Ce $f$-shell from LDA, GGA and LDA+SO calculations.}
\label {fig:lda} 
\end {figure}

DFT+DMFT method is used recently to investigate wide class of strongly correlated materials and allows to understated microscopic nature of magnetic, electronic and structure phase transition in 3$d$ metals 
 The calculation scheme is constructed in the following way: first, a Hamiltonian is produced using converged DFT results for the system under investigation, then the many-body  Hamiltonian is set up, and finally the corresponding self-consistent DMFT equations are solved.

In practical calculations, $U$ is often considered as a free parameter to achieve the best agreement of calculated and measured properties of investigated system. Alternatively, $U$ value could be estimated from the experimental spectra. However, the most attractive approach is to determine Coulomb interaction parameter $U$ value in \textit {first principles} non-empirical way. There are two such methods: constrained DFT scheme~\cite {U-calc,anigun}, where  $d$-orbital occupancies in DFT calculations are fixed to the certain values and $U$ is numerically determined as a derivative of $d$-orbital energy over its occupancy, and  Random Phase Approximation (RPA) method~\cite {RPA}, where screened Coulomb interaction between $d$-electrons is calculated via perturbation theory.

Ab-initio calculations of electronic structure  were   obtained within TB-LMTO-ASA ~\cite{LMTO} and the pseudopotential plane-wave method PWSCF, as implemented in the Quantum ESPRESSO package~\cite{PW}. In the latter scalar-relativistic PBE pseudopotential was used.  The Hamiltonians $\hat H_{DFT}$ in Wannier function (WF) basis~\cite{Wannier37, MarzariVanderbilt} were produced using projection procedure  that is  described in details in Ref.~\cite{Korotin}. In order to include SOC in DFT+DMFT scheme LDA+SO calculations were carried out as described in details elsewhere~\cite{Shorikov05}. This method was used to describe spectral properties of Pu.~\cite{Anisimov07}
        
The basis set of used Hamiltonians includes all bands  that are formed by $s$-, $p$-, $d$-  and $f$-Ce states. That would correspond to the extended model where in addition to correlated $f$-orbitals all $s,p,d$-orbitals are included too. 

The resulting Hamiltonian to be solved by DMFT has the form:
\begin{equation}
\hat H= \hat H_{DFT}- \hat H_{dc}+\frac{1}{2}\sum_{i,\alpha,\beta,\sigma,\sigma^{\prime}}
U^{\sigma\sigma^{\prime}}_{\alpha\beta}\hat n^{f}_{i\alpha\sigma}\hat n^{f}_{i\beta\sigma^{\prime}},
\label{eq:ham}
\end{equation}
where $U^{\sigma\sigma^{\prime}}_{\alpha\beta}$ is the Coulomb interaction matrix, 
$\hat n^d_{i\alpha\sigma}$ is the occupation number operator 
for the $f$ electrons with orbitals $\alpha$ or $\beta$ and spin indices $\sigma$ 
or $\sigma^{\prime}$ on the $i$-th site. 
The term $\hat H_{dc}$ stands for the {\it d}-{\it d} interaction 
already accounted for in LDA, so called double-counting correction. 
In the present calculation the double-counting was chosen in 
the following form $\hat H_{dc}=\bar{U}(n_{\rm dmft}-\frac{1}{2})\hat{I}$.
Here $n_{\rm dmft}$ is the self-consistent total number of {\it f} electrons 
obtained within the GGA+DMFT, $\bar{U}$ is the average Coulomb parameter for 
the {\it d} shell and $\hat I$ is unit operator. 

The on-site Coulomb repulsion parameter ($U$) was set to be 6.0 eV as in previous works \cite{Zolfl01, Held01}. The intra-atomic Hund's rule coupling was set to $J_H$=0 eV in several calculation or to value 0.57 eV calculated in frames of constrained LDA method.~\cite{U-calc}
 
The effective  impurity problem for the DMFT was solved by the hybridization  expansion Continuous-Time Quantum Monte-Carlo method (CT-QMC) \cite{CTQMC}.  Calculations for all volumes were performed in the paramagnetic state at the inverse temperature $\beta=1/T$ = 10
eV$^{-1}$  corresponding to 1160~K.
Spectral functions on real energies  were calculated by Maximum Entropy Method (MEM).~\cite{mem}

\begin {figure}
\vspace{1cm}
\includegraphics [width=0.45\textwidth]{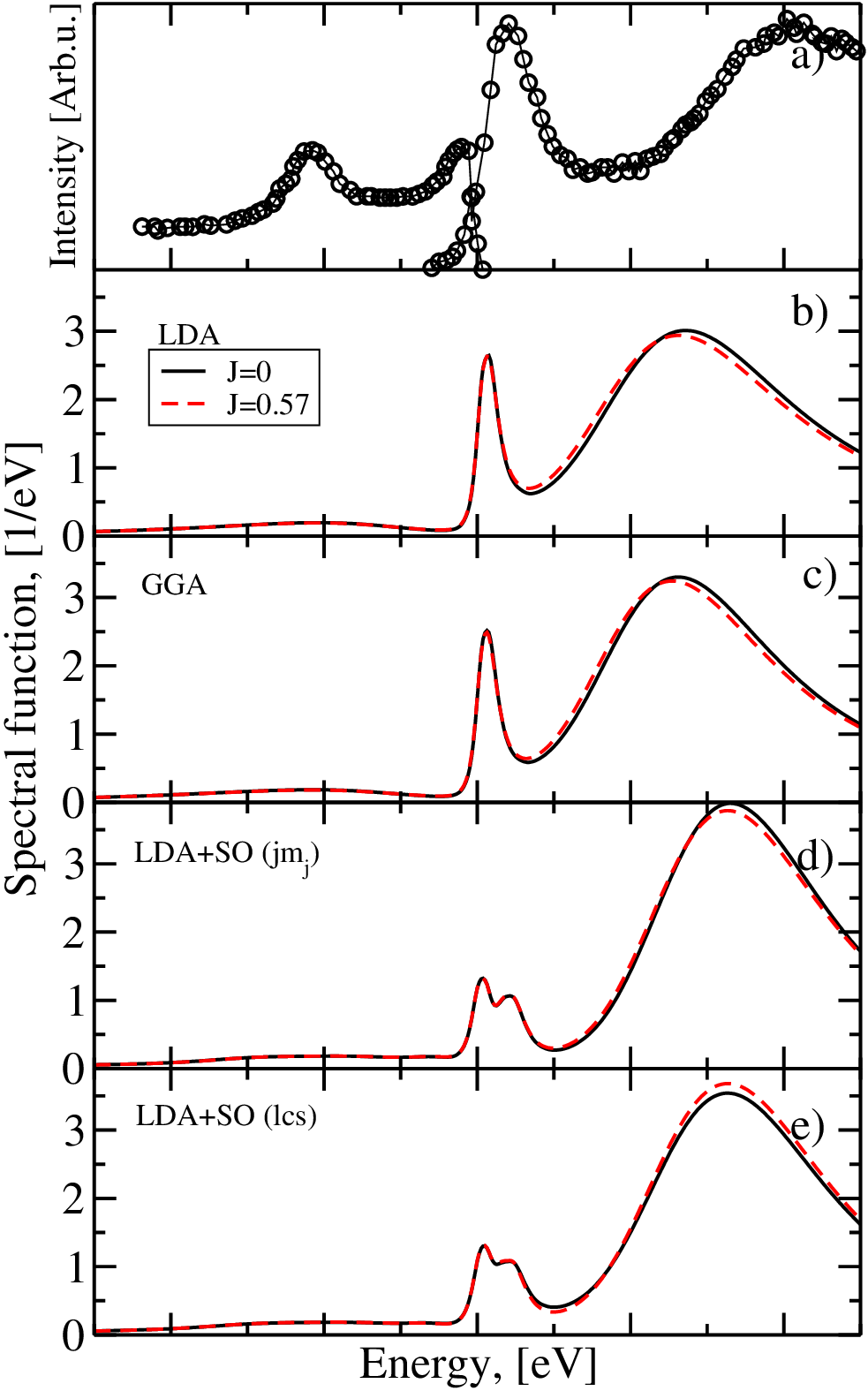}
\caption {Comparison between combined PES and BIS spectra (panel a) and DFT+DMFT spectral function of $\alpha$-Ce   
for $\beta$ = 10 eV$^{-1}$. Hamiltonian was obtained by LDA (TB+LMTO+ASA) (panel b), GGA (QE) (c), LDA+SO ($jm_j$ basis) (d) and LDA+SO LCS basis (e). Coulomb parameters were U=6 eV, J=0 eV (black solid lines) and U=6 eV J=0.57 eV (red dashed lines).}
\label {fig:alpha} 
\end {figure}

Since  on-site correlations were accounted for in density-density one should take care if diagonal elements of Green function matrix  (and hence Hamiltonian matrix) are small and made diagonalization procedure if necessary.  
Full moment basis $jm_j$ is natural one if SOC is strong. If intermediate coupling scheme is appropriate due to competition between SO and Hund's exchange Hamiltonian matrix in $jm_j$ basis become sparse one and should be additionally diagonalized due to restriction of DMFT solver.  
With accordance to crystal field theory, in the cubic symmetry j = 5/2 states separate into $\Gamma_7$ doublet and  $\Gamma_8$ quartet. Diagonalization of occupation matrix reveal this behavior. Then found eigenvector matrices were used to diagonalize LDA+SO Hamiltonian. Both bases $jm_j$ and local coordinate system (denoted LCS hereafter) were used in this work.

\begin {figure}
\vspace{1cm}
\includegraphics [width=0.5\textwidth]{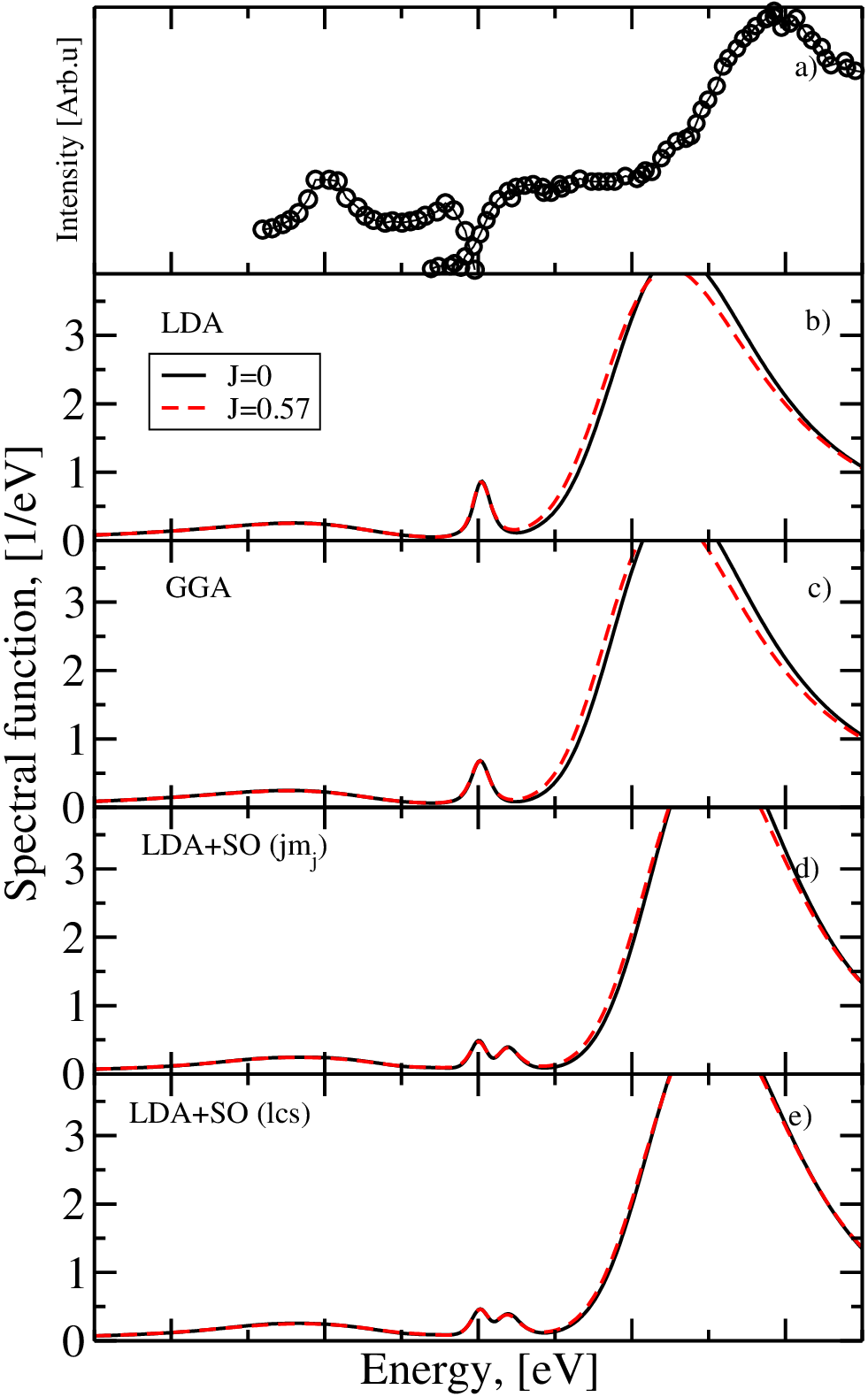}
\caption {Comparison between combined PES and BIS spectra (panel a) and DFT+DMFT spectral function of $\gamma$-Ce   
for $\beta$ = 10 eV$^{-1}$. Hamiltonian was obtained by LDA (TB+LMTO+ASA) (panel b), GGA (QE) (c), LDA+SO ($jm_j$ basis) (d) and LDA+SO LCS basis (e). Coulomb parameters were U=6 eV, J=0 eV (black solid lines) and U=6 eV J=0.57 eV (red dashed lines).}
\label {fig:gamma} 
\end {figure}

Obtained DOS form LDA, GGA and LDA+SO calculation are shown in the Fig~\ref{fig:lda}. First two results are almost similar and no difference in DFT+DMFT results is expected. SOC splits narrow Ce f-shell,  and should change DFT+DMFT spectral function. 

Results of DFT+DMFT calculations are presented in Fig.~\ref{fig:alpha}-\ref{fig:gamma}. In all obtained spectra both upper and lower Hubbard bands are well distinguished. The interval between their centers corresponds to applied $<U>$ = 6 eV. The pronounced quasiparticle peak is seen near the Fermi level (in all the figures correspond to 0 eV). Small difference in band structure obtained by LDA and GGA methods are smoothed in DFT+DMFT calculations and results on panels a) and b) are almost the same. Switching on of Hund exchange don't change the picture as would be expected since Ce has only one electron in f-shell and it has nothing to interact with.

On the other hand taking into account SOC leads to drastic change in the  feature near Fermi level though upper and lower Hubbard bands remain nearly unchanged. In contrast the quasiparticle peak splits into two originated from splitting of LDA f-band into j=5/2 and j=7/2 subbands. Correlation effects renormalize LDA band structure it become narrow but keep main features. Obtained shape of spectral function near Fermi level agrees well with the peculiarities in PES and BIS spectra.~\cite{pesbis} Also, in this case including of Hund exchange interaction doesn't change obtained spectral function. 

In conclusion, we have investigated the importance of Hund's interaction and SOC for Ce spectral function calculation. SOC results in the splitting of quasiparticle peak which originate from 5/2 7/2 sub-bands. Accounting of Hund's interaction doesn't play an important role since Ce have only one electron in correlated f-shell.

This work were supported by the grant of the Russian Scientific Foundation (project no. 14-22-00004).

\end {document}